\documentclass[twocolumn,twoside,showkeys,showpacs,preprintnumbers,superscriptaddress,amsmath,amssymb,prc]{revtex4-2}

\usepackage{graphicx}
\usepackage{dcolumn}
\usepackage{bm}
\usepackage[version=4]{mhchem}
\usepackage{subcaption}
\usepackage[colorlinks=true,linkcolor=black,citecolor=blue,urlcolor=blue,]{hyperref}
\usepackage{appendix}
\usepackage{ulem} 
\usepackage[justification=raggedright,singlelinecheck=false]{caption}
\usepackage{booktabs}
\usepackage{multirow}
\usepackage{hyperref}

\begin{document}

\title{Linear Response of CsI(Tl) Crystal to Energetic Photons below 20 MeV}

\author{Junhuai Xu}
\email{xjh22@mails.tsinghua.edu.cn}
\affiliation{Department of Physics, Tsinghua University, Beijing 100084, China}
\author{Dawei Si}
\email{sdw21@mails.tsinghua.edu.cn}
\affiliation{Department of Physics, Tsinghua University, Beijing 100084, China}
\author{Yuhao Qin}%
\affiliation{Department of Physics, Tsinghua University, Beijing 100084, China}
\author{Mengke Xu}%
\affiliation{Shanghai Advanced Research Institute, Chinese Academy of Sciences, Shanghai 201210, China}
\author{Kaijie Chen}%
\affiliation{Shanghai Advanced Research Institute, Chinese Academy of Sciences, Shanghai 201210, China}
\author{Zirui Hao}%
\affiliation{Shanghai Advanced Research Institute, Chinese Academy of Sciences, Shanghai 201210, China}
\author{Gongtao Fan}%
\affiliation{Shanghai Advanced Research Institute, Chinese Academy of Sciences, Shanghai 201210, China}
\author{Hongwei Wang}%
\affiliation{Shanghai Advanced Research Institute, Chinese Academy of Sciences, Shanghai 201210, China}
\author{Yijie Wang}
\affiliation{Department of Physics, Tsinghua University, Beijing 100084, China}
\author{Zhigang Xiao}
\email{xiaozg@tsinghua.edu.cn}
\affiliation{Department of Physics, Tsinghua University, Beijing 100084, China}
\affiliation{Center of High Energy Physics, Tsinghua University, Beijing 100084, China}

\date{\today}

\begin{abstract}
 
The linear response of CsI(Tl) crystals to $\gamma$-rays plays a crucial role in their calibration, as any deviation from linearity can introduce  systematic errors not negligible  in the measurement of $\gamma$ energy spectra, particularly at high energies. In this study, the responses of CsI(Tl) crystals to high-energy photons up to 20 MeV are investigated using quasi monochromatic  $\gamma$ beam provided by the Shanghai Laser Electron Gamma Source. The spectra are folded using a detector filter implemented by Geant4. Both quadratic and linear fits to six energy points are used to assess the linearity of the CsI(Tl) detector. The results demonstrate that the difference between the linear and non-linear fits is at the level of 4\%. Applying these findings to the $\gamma$ hodoscope of the Compact Spectrometer for Heavy Ion Experiment (CSHINE), the potential systematic uncertainties caused by CsI(Tl) non-linearity are evaluated. This work provides a comprehensive calibration methodology for employing  CsI(Tl) crystal to detect high energy $\gamma$-rays.

\end{abstract}

\maketitle

\section{Introduction}

CsI(Tl) crystals are widely utilized in spectrometers across various energy ranges due to their high luminosity, density, and exceptional efficiency and resolution for detecting high-energy photons. Notable applications in particle physics include for examples the electromagnetic calorimeters in BES-III \cite{BESIII-const} and BaBar \cite{BaBar-EMC}. More recently, the Compact Spectrometer for Heavy Ion Experiment (CSHINE) has incorporated a CsI(Tl) hodoscope to measure the bremsstrahlung $\gamma$ energy spectrum in heavy-ion reactions \cite{Qin:2022mzp}.

Accurate calibration is essential for the optimal performance of spectrometers. The CSHINE $\gamma$ hodoscope uses \ce{^{60}Co} and natural \ce{ThO_2} sources to calibrate the energy response of CsI(Tl) crystals \cite{Qin:2022mzp}. However, these sources emit photons only up to approximately 2.6 MeV, necessitating an extrapolation by nearly a factor of 20 for high-energy measurements. Consequently, understanding the linearity of the CsI(Tl) response is critical for precise reconstruction of high-energy $\gamma$ spectra. Previous studies employing proton capture reactions, such as \ce{^{19}F(p, \alpha\gamma)^{16}O} and \ce{^{7}Li(p, \gamma)^{8}Be}, suggest that the CsI(Tl) response remains linear within about 2-4\% \cite{Qin:2022mzp}. However, these methods involve thick compound targets, which result in complex spectral features that complicate accurate quantification of non-linearity.

The newly commissioned Shanghai Laser Electron Gamma Source (SLEGS) serves as a beamline station in the Phase II Project of the Shanghai Synchrotron Radiation Facility (SSRF). It provides tunable $\gamma$-ray beams with continuous energy coverage ranging from 0.25 to 21.1 MeV, produced via Compton scattering of laser photons with electrons. By backscattering, the $\gamma$ energy is 21.7 MeV \cite{SLEGS-comm}. The clean, quasi monochromatic $\gamma$-ray spectra generated at SLEGS provide an ideal platform for studying the non-linearity of the CsI(Tl) energy response with high precision.

In this paper, we study the linear response of the CsI(Tl) to the $\gamma$-rays up to about 20 MeV, using the quasi monochromatic $\gamma$-rays provided by SLEGS. The remainder of the paper is organized as following. Section II describes the experimental setup. Section III reports the results of the experiment, focusing mainly on the non-linearity of the CsI(Tl). Section IV demonstrate the method to conduct the calibration of the CsI(Tl) for measuring the bremsstrahlung $\gamma$-rays in heavy ion reactions done at CSHINE. Systematic uncertainty has been evaluated. Section V is the conclusion.

\section{Experimental Setup and Measurements}

\subsection{Facility and $\gamma$ beam}
The calibration experiment was conducted at the SLEGS beamline of SSRF. The primary components of the SLEGS beamline responsible for $\gamma$-ray generation include the interaction chamber, the multi-functional chamber and a $\rm CO_2$ laser \cite{XU2022166742}. The laser can generate backscattering $\gamma$-rays with energy of 21.7 MeV. In slant scattering mode in the interaction chamber,  $\gamma$-rays are produced with variable energies  via Compton scattering. By adjusting the scattering angle $\theta_{\rm e \text -\gamma}$ between the laser and the electron from $20^\circ \text - 160^\circ$, the $\gamma$ energy can be continuously tuned \cite{Liu:2024eks}.

\begin{figure}[hbtp]
    \centering
    \includegraphics[width=0.45\textwidth]{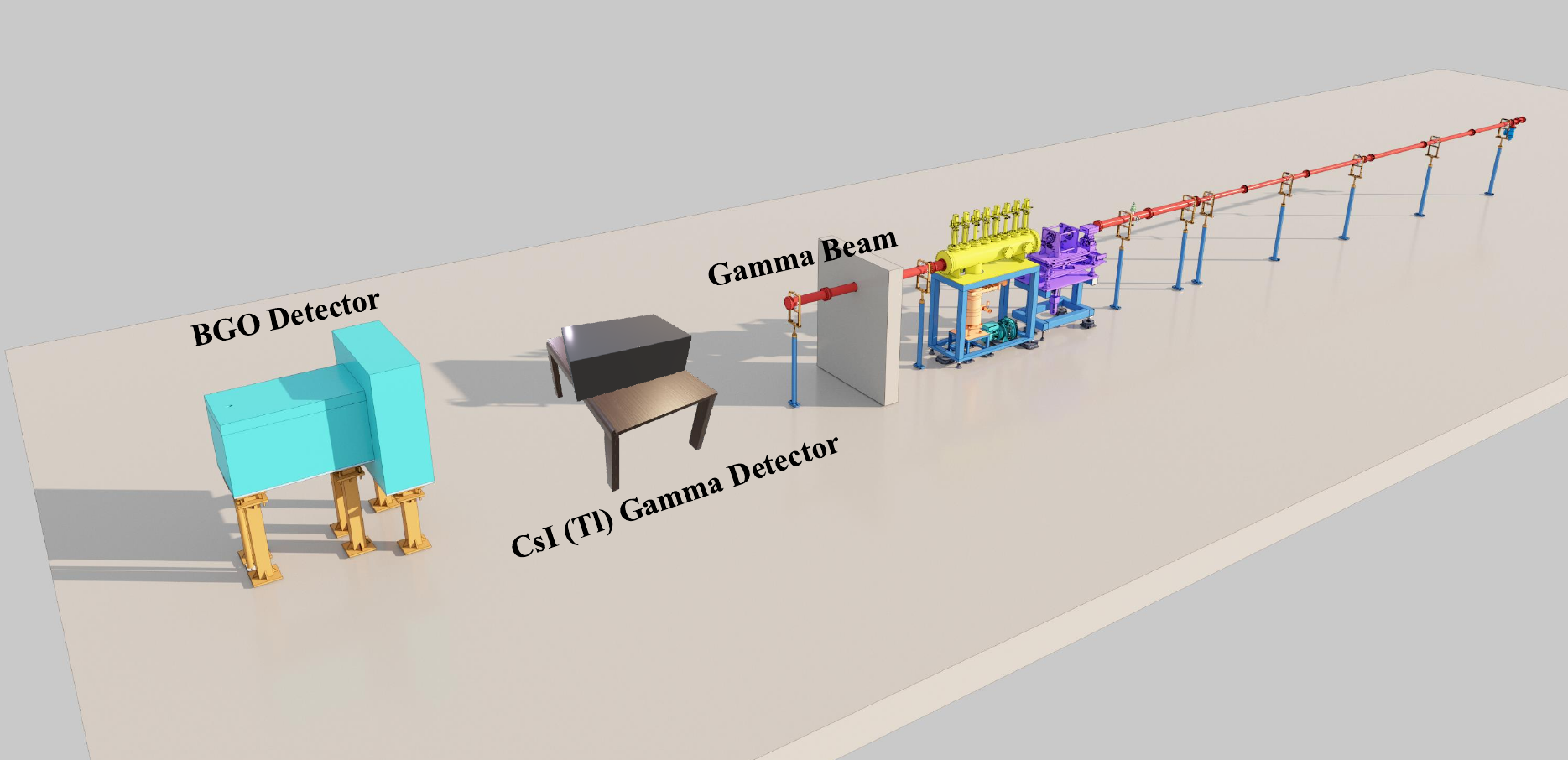}
    \caption{(Color Online) Schematic of the SLEGS beamline experimental setup. The $\gamma$ beam is emitted from the red pipeline, with the BGO detector at the end serving as a reference for standard measurements.}
    \label{SLEGS}
\end{figure}

The schematic view of the experimental setup at the SLEGS beamline is shown in Fig.~\ref{SLEGS}. The $\gamma$ beam is emitted from the red-colored pipeline and measured by the BGO detector (blue box), which is positioned at the end of the beamline as a beam monitor.  The BGO detector has been thoroughly calibrated and characterized. Hence it can be used to obtain  the energy response spectrum of the $\gamma$ beam and further to deduce the original energy spectrum by unfolding the detector filter effect \cite{Liu:2024mjr}. Specifically, the energy resolution of BGO detector at different $\gamma$-ray energies was calibrated using monoenergetic $\gamma$-rays produced in $(p,\gamma)$ reactions, given by $\Delta E/E=20.600/E+3.436\times\sqrt{1/E+6.664\times10^{-5}}$  with $E$ in units of keV. By incorporating this resolution function into Geant4 simulations, the response functions of the detector can be determined. These response functions are then used in an iterative least-squares fitting procedure to extract the unfolded $\gamma$-ray spectrum from the total-energy response of the BGO detector \cite{Liu:2024mjr}. For each beam setting, the BGO detector is first used to measure the beam properties. Then the CsI detector being studied is placed on the test bench between the $\gamma$ beam exit window and the BGO detector to conduct the calibration experiment.

\begin{table}[hbtp]
    \centering
    \begin{tabular}{cccccc}
    \toprule
    Beam Index  &   $\theta_{\rm e \text -\gamma}$  & $\hat E_\gamma^{\rm BGO}$ (MeV) & $\hat E_\gamma^{\rm Ori}$ (MeV) & $\delta E_\gamma$\\ 
    \midrule 
    1 & $60^{\circ}$  & 4.9(3)     & 4.9(3)  &15.3\% \\
    2 & $75^{\circ}$  & 7.4(4)     & 7.6(4)  &12.8\% \\
    3 & $90^{\circ}$  & 10.1(5)    & 10.2(5) &9.6\% \\
    4 & $104^{\circ}$ & 12.9(6)    & 13.0(6) &8.1\% \\
    5 & $120^{\circ}$ & 15.8(6)    & 15.9(6) &6.0\% \\
    6 & $130^{\circ}$ & 17.5(7)    & 17.6(7) &5.0\% \\
    \bottomrule
    \end{tabular}
    \caption{Beam information including $\theta_{\rm e \text -\gamma}$, peak positions of the measured spectra $E_\gamma^{\rm BGO}$ and the unfolded spectra $E_\gamma^{\rm Ori}$, along with the energy dispersion of the latter for each beam index.}
    \label{Beaminfo}
\end{table}

\begin{figure}[hbtp]
    \centering
    \includegraphics[width=0.45\textwidth]{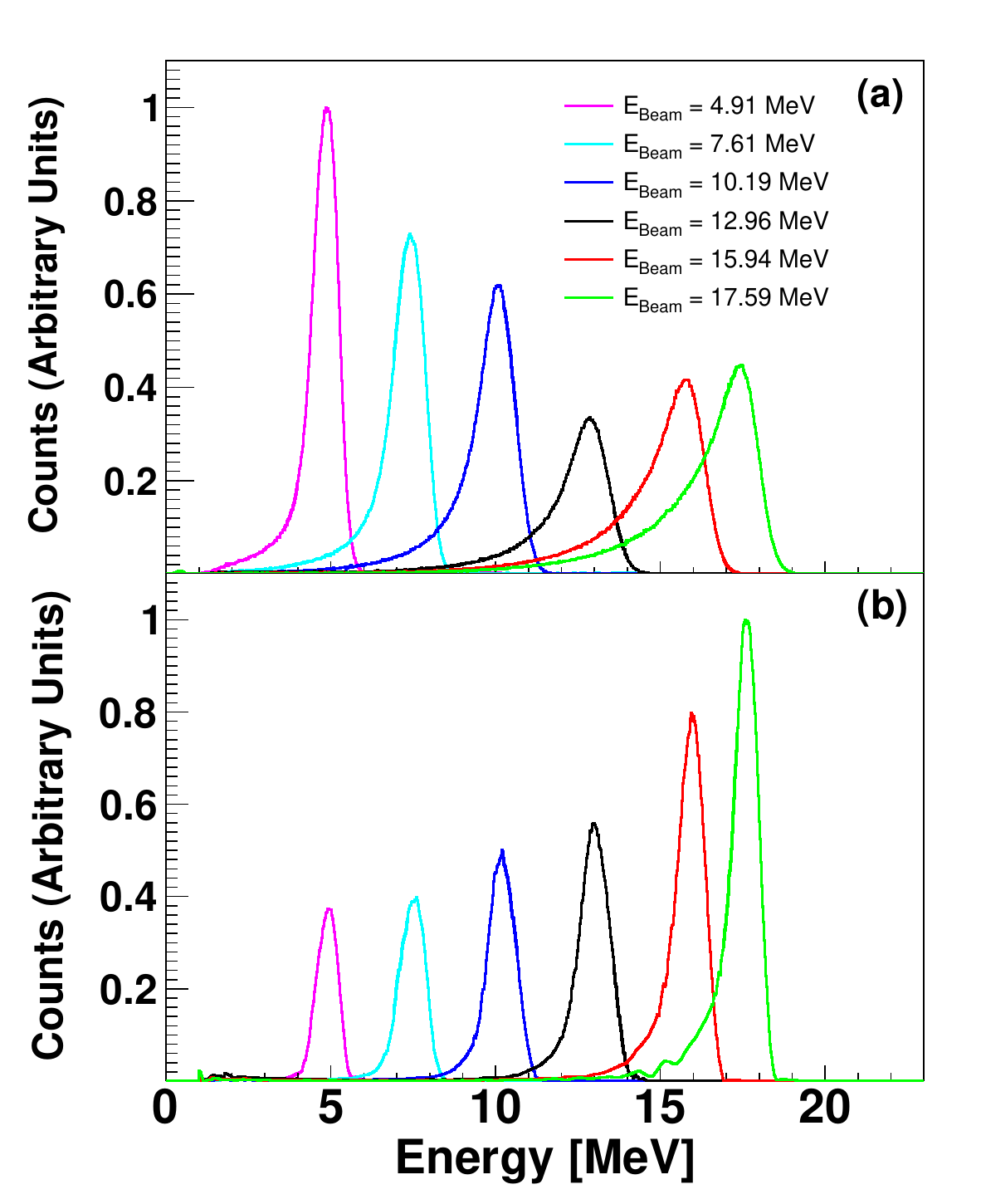}
    \caption{(Color Online) $\gamma$ spectra measured by the BGO detector. (a) The raw spectra directly measured by the BGO detector at various beam settings, represented by different colors. (b) The original $\gamma$ spectra obtained through deconvolution using the unfolding algorithm.}
    \label{EnergySpectraBGO}
\end{figure}

During the calibration of the CsI(Tl) crystal detectors, six $\gamma$ energy points were used, with the corresponding scattering angles for each beam as listed in Table~\ref{Beaminfo}. The energy response spectra measured by the BGO detector, $E_\gamma^{\rm BGO}$, for the six beams are presented in Fig.\ref{EnergySpectraBGO}(a), while the  original energy spectra $E_\gamma^{\rm Ori}$ by unfolding the detector filter effect are shown in Fig.\ref{EnergySpectraBGO}(b). An approximate constant counting rate of the CsI(Tl) $\gamma$ detector is achieved by adjusting the attenuation rate of beam at each setting. All spectra in each panel have been normalized by their respective highest peak values to facilitate better visualization. Because of the leakage of the energy and the influence of finite energy resolution, the measured spectra is slightly broad and  shifted to the left side. The peak positions of the measured and the unfolded $\gamma$ energy spectra  can be extracted. The corresponding values are summarized in Table~\ref{Beaminfo}. The uncertainty of each peak energy is derived from the inherent resolution of the BGO detector. In addition, the energy dispersion of photopeak $\delta E_\gamma$ for the six $\gamma$-ray beams are also presented in Table~\ref{Beaminfo}.

\subsection{Detector Design}

The CsI(Tl) crystals employed in the measurement of the bremstrahlung $\gamma$-rays in heavy ion reactions \cite{Qin:2022mzp}, and to be tested here, are produced by the Institute of Modern Physics, Chinese Academy of Sciences \cite{ChenRuo-Fu_2008,SHI2020835}.  The CsI(Tl) scintillator, a high-Z material, is chosen for its low deliquescence and ease of machining. The radiation length of CsI(Tl) is $X_0 = 8.39~ \rm g/cm^2$, and its Molière radius is $R_{\rm M} = 3.531~ \rm cm$.  Each crystal unit has dimensions of $70 ~\rm mm \times 70 ~mm \times 250~ mm$. One of the $70 ~\rm mm \times 70~ mm$ surfaces serves as the $\gamma$-ray entrance, corresponding to a transverse dimension of approximately $1.98~ R_{\rm M}$, while the $250 ~ \rm mm$ length corresponds to roughly $13.4 ~X_0$.

The CsI(Tl) scintillator crystal is wrapped in Teflon tape to enhance light reflection. The unwrapped rear surface  is coupled to a Hamamatsu R6231 photomultiplier tube (PMT) using BC-631 silicone grease from Saint-Gobain Crystals. To prevent light leakage, the wrapped crystal and PMT assembly are further enclosed using black tape to enhance the light shielding. The PMT is connected to a Hamamatsu E1198-23 socket, which distributes the high voltage supplied by a CAEN NDT1470 module and extracts the signal via AC coupling. The entire detector unit is then mounted in a 3D-printed  shell to ensure the structural stability.

\subsection{Electronics System}

\begin{figure}[hbtp]
    \centering
    \includegraphics[width=0.5\textwidth]{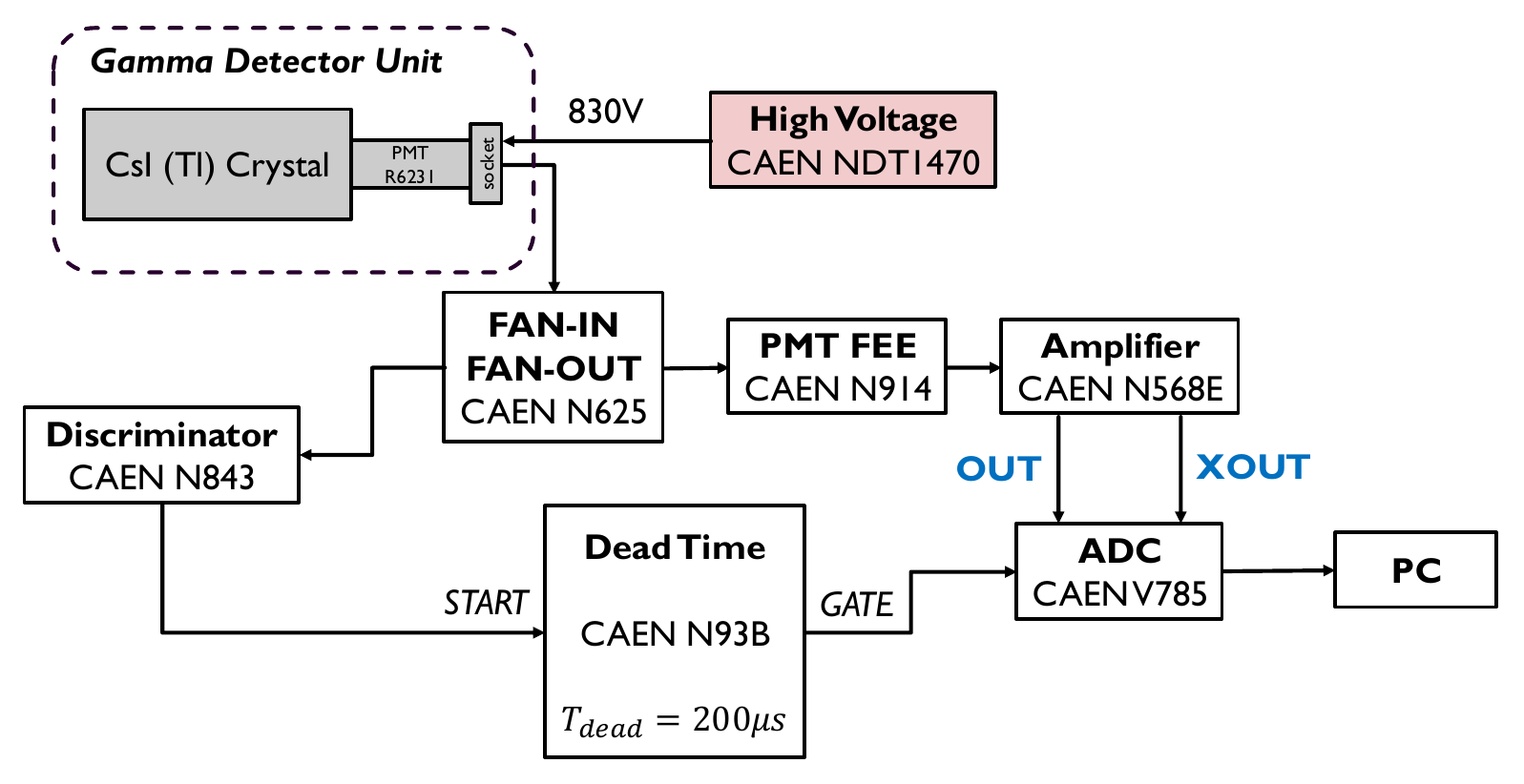}
    \caption{(Color Online) Layout of the electronic and data acquisition system.}
    \label{DAQLayout}
\end{figure}

The electronics of the $\gamma$  detector consist of a high voltage module, front-end electronics (FEE), a spectrometer amplifier, a peak-sensing analog-to-digital converter (ADC), and a data acquisition (DAQ) system, as Fig.~\ref{DAQLayout} shown. The high voltage module supplies $830\rm V$ to the PMT, and the resulting signal is transmitted to the FEE module, CAEN N914, which delivers the amplitude signal. The negatively polarized amplitude signals from the N914 are then sent to the CAEN N568E spectrometer amplifier. To record both high- and low-energy $\gamma$-rays, the output signals from the OUT (ADC-E) and XOUT (ADC-XE) channels of the N568E, with gain factors of 8 and 80, respectively, are fed into the peak-sensing ADC, CAEN V785. The low-gain output (OUT) is primarily used for detecting high-energy $\gamma$-rays, while the high-gain output (XOUT) is used for calibrations with low-energy radioactive sources. To prevent the DAQ system from entering an unstable state due to multiple adjacent trigger signals, a $200\rm \mu s$ dead time is applied after each trigger signal input by the CAEN N93B.

\begin{figure*}[hbtp]
    \centering
    \includegraphics[width=0.8\textwidth]{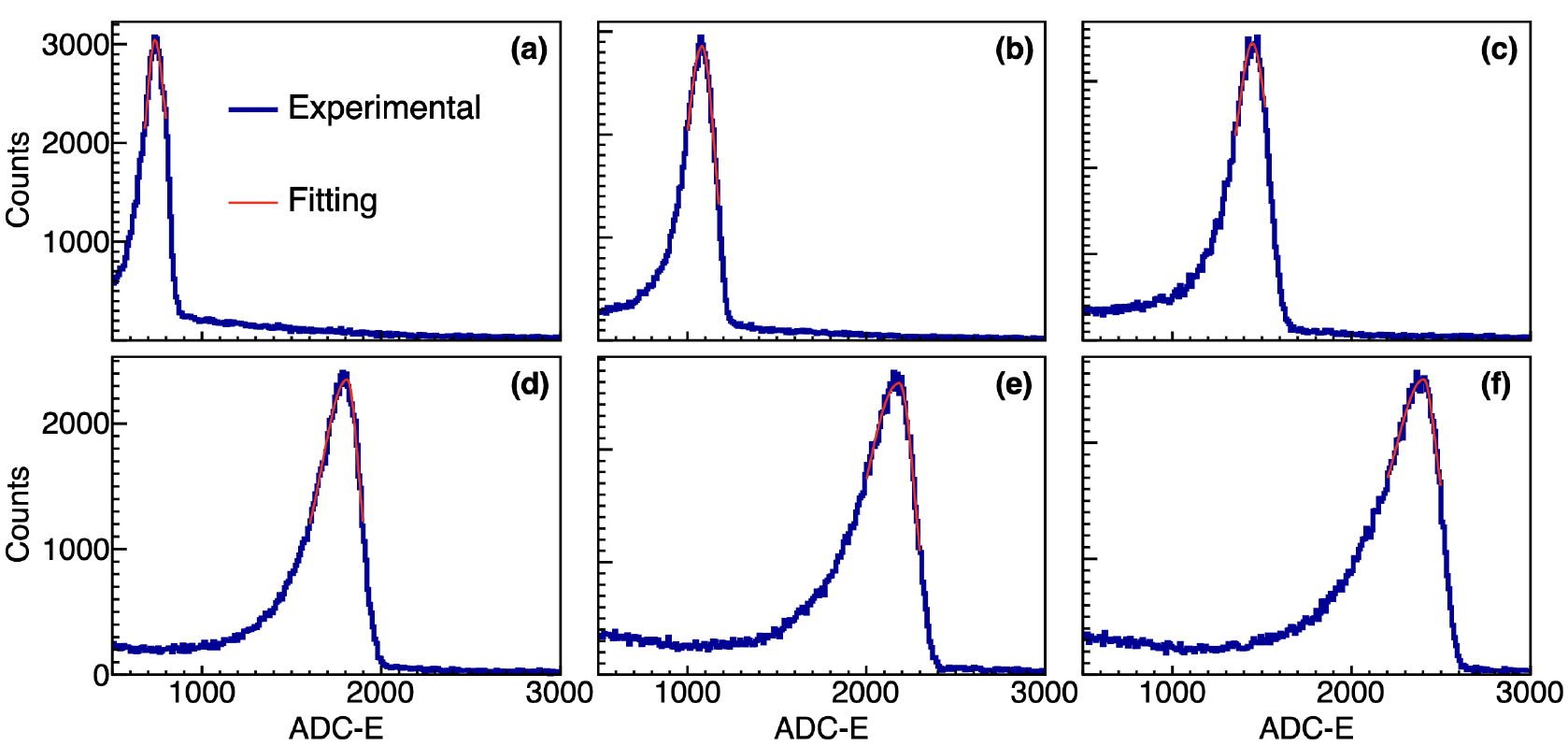}
    \caption{(Color Online) $\gamma$ energy spectra measured by the selected CsI(Tl) detector (Det04) for different $\gamma$ beams and their corresponding asymmetric Gaussian fits. Panels (a)-(f) show spectra for different $\gamma$ beam indices, with the blue dashed curves representing the measured spectra and the red solid lines indicating the results of the asymmetric Gaussian fit.}
    \label{expHistos}
\end{figure*}

\section{Results and Discussions}

\subsection{Raw $\gamma$ energy spectra measurement}

Fig.~\ref{expHistos} presents the measured energy spectra of one CsI(Tl) detector, taking Det04 for  example. For the high energy $\gamma$-rays, the output of the ADC-E channel is analyzed. It is shown the peaks exhibit a left-right asymmetry in accordance with the  BGO calibration detector. To better capture the spectral information, an asymmetric Gaussian function was chosen to fit the peak region of the spectra. The asymmetric Gaussian function is described by Eq.~(\ref{asymGausfunc}),
\begin{equation}\label{asymGausfunc}
    f(x)=
    \begin{cases}
    \frac{2A}{\sqrt{\pi(\beta_l+\beta_r)}}e^{-(\frac{x-\mu}{\beta_l})^2}+C, &\quad x\le\mu\\
    \frac{2A}{\sqrt{\pi(\beta_l+\beta_r)}}e^{-(\frac{x-\mu}{\beta_r})^2}+C, &\quad x>\mu\\
    \end{cases}
\end{equation}
where $A$ represents the amplitude, $\mu$ denotes the peak position, $C$ is the baseline, and $\beta_l$ and $\beta_r$ are the widths of left and right half of the peak, respectively. The fitting results are shown by the red solid curves in Fig.~\ref{expHistos}. Based on the fit, the peak positions of the spectra for each beam energy were extracted, along with the corresponding ADC-E values ($E_a$). This data can be used to calibrate the detector by comparing the fitted peak positions with the corresponding energy values (see next).

\begin{figure*}[hbtp]
    \centering
    \includegraphics[width=0.8\textwidth]{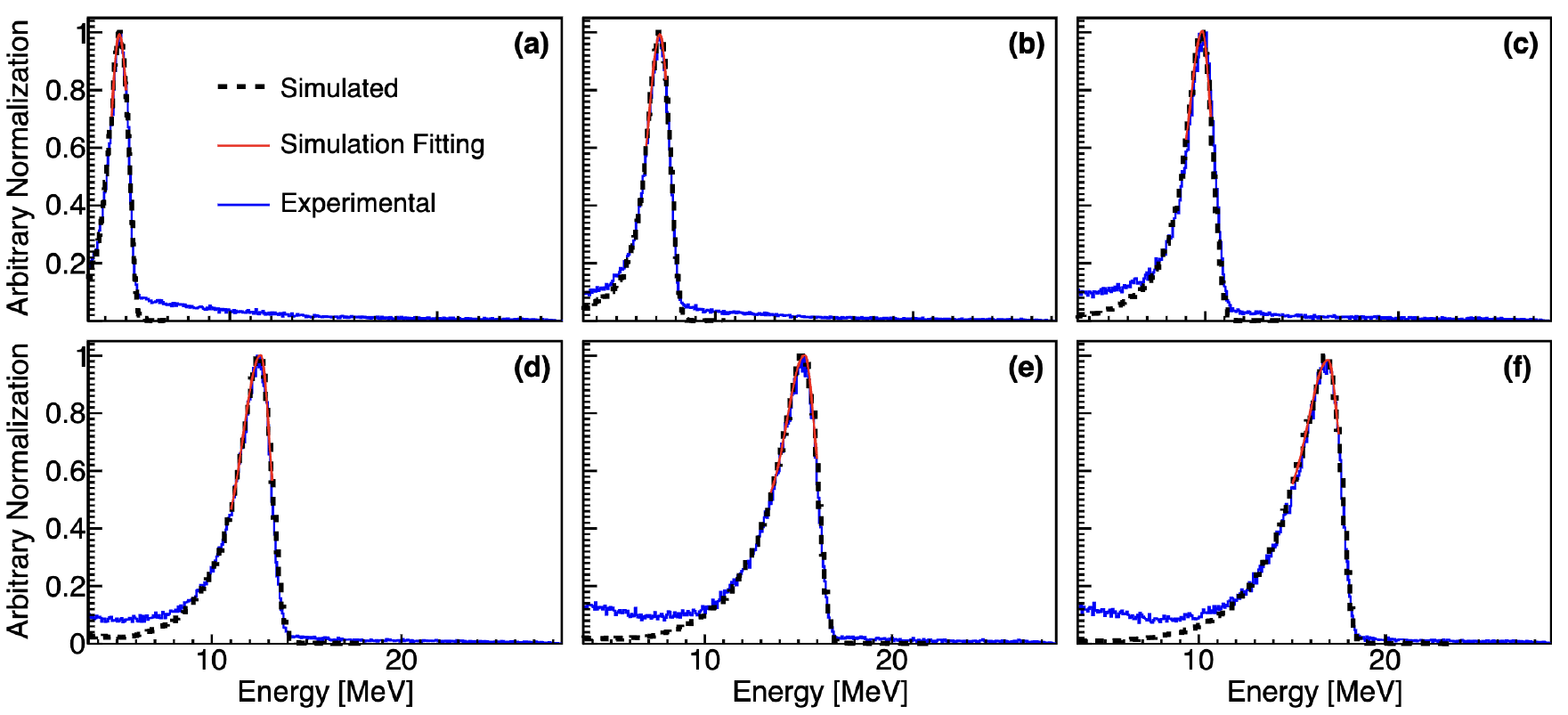}
    \caption{(Color Online) Comparison between simulated and experimental spectra of the selected CsI detector (Det04). Panels (a)-(f) correspond to six different beam energies. The black dashed lines and blue solid lines represent the normalized simulated and experimental spectra, respectively, with the experimental spectra aligned to the x-axis according to their peak positions in comparison to the simulated spectra. The red solid curves represent the asymmetric Gaussian fit to the peaks of the simulated spectra.}
    \label{SimHistos}
\end{figure*}

\subsection{Response Simulations}

Similar to the BGO calibration detector, the measured spectra are influenced by energy leakage and finite resolution, necessitating the simulation of CsI(Tl) response spectra to the original ones. Geant4 simulation packages were used for this purpose. A CsI crystal with dimensions of $70 ~ \rm mm \times 70 ~ mm \times 250 ~ mm$ was modeled using the ``G4\_CESIUM\_IODIDE" material in Geant4 and placed in air (``G4\_AIR"). For a given $\gamma$  beam energy, the incident energy of each event was sampled from the unfolded spectrum shown in Fig.~\ref{EnergySpectraBGO}(b), and a 1\% detector energy resolution effect was applied (using Det04 as an example). This resolution setting, which accounts for slight variations among different CsI detectors due to manufacturing tolerances and assembly conditions, was introduced to ensure consistency between the simulated and experimental response spectra. The simulated response spectra are shown as black dashed lines in panels (a)-(f) of Fig.~\ref{SimHistos}, with each spectrum normalized to its peak height. Again the asymmetric Gaussian fittings were applied to the peaks, represented by the red solid curves. The fitted peak positions correspond to the response energy  $E_s$  of the CsI detector for each beam energy. The correspondence between  $E_s$  and the unique ADC-E value  $E_a$  from the experimental spectra enables proportional scaling of the latter, aligning their peak positions with those of the simulated ones, as shown by the blue solid lines in panels (a)-(f) of Fig.~\ref{SimHistos}. The close agreement between the simulated and experimental spectra demonstrates the accuracy of the simulation. The same procedure was applied to the other CsI detectors, where the energy resolution parameters were individually optimized for each detector and then used in the subsequent analysis. The selected energy resolutions in the simulation were 3\%, 1.2\%, and 2.5\% for Det01, Det02, and Det03, respectively.

\begin{figure}[hbtp]
    \centering
    \includegraphics[width=0.5\textwidth]{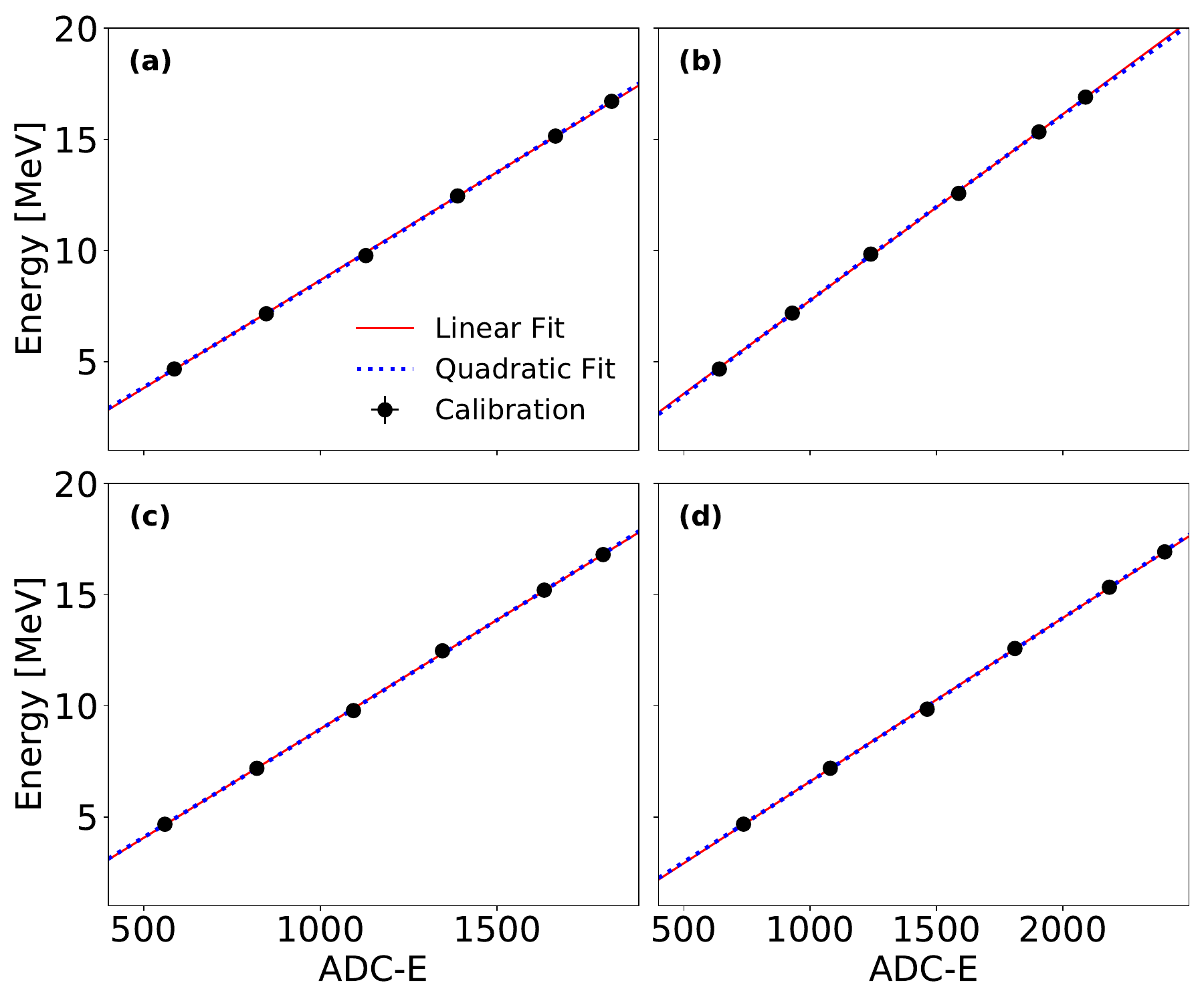}
    \caption{(Color Online) Calibration and fit results for CsI $\gamma$ detectors. The black points represent the calibration results from both simulated and experimental spectra, with the red solid line indicating the linear fit and the blue dashed line showing the quadratic fit.}
    \label{LinearResults}
\end{figure}

\begin{table*}[hbtp]
    \centering
    \begin{tabular}{c|cc|cccc|cc}
    \toprule
    \multirow{2}{*}{CsI Index}  &  \multicolumn{2}{c|}{Linear ($E_{\gamma}=l+k\text{CH}_{\text{E}}$)}  & \multicolumn{4}{c|}{Quadratic ($E_{\gamma}=p_0+p_1\text{CH}_{\text{E}}+p_2\text{CH}_{\text{E}}^2$)} &\multirow{2}{*}{$|\frac{\epsilon_2}{p_2}|$} &\multirow{2}{*}{$|\frac{p_2}{k}\text{CH}_m|$} \\ 
      &$l$ &$k$ ($\times10^{-3}$) &$p_0$ &$p_1$ ($\times10^{-3}$) &$p_2$ ($\times10^{-7}$) &$\epsilon_2$ ($\times10^{-7}$) \\
    \midrule 
    Det01 & -1.06(3)  & 9.73(2)   & -0.85(8)  &9.27(17) &2.03    &0.74  &0.37 &0.045 \\
    Det02 & -0.64(4)  & 8.39(3)   & -0.85(11) &8.75(17) &-1.39   &0.65  &0.46 &0.041 \\
    Det03 & -0.86(4)  & 9.82(4)   & -0.72(13) &9.54(26) &1.22    &1.15  &0.94 &0.026 \\
    Det04 & -0.77(4)  & 7.36(2)   & -0.60(11) &7.10(16) &0.87    &0.53  &0.61 &0.033 \\
    \bottomrule
    \end{tabular}
    \caption{Fit results for CsI(Tl) $\gamma$ detectors. Parameters obtained from linear and quadratic fits for four different CsI(Tl) crystal detectors. To assess the degree of non-linearity in the energy response, we calculate (i) the relative uncertainty of the quadratic term coefficient and (ii) the relative magnitude of the quadratic term coefficient compared to the linear term coefficient, scaled by the ADC-E value corresponding to 20 MeV in the linear fit of each CsI detector.}
    \label{FittingCompare}
\end{table*}

\subsection{Linearity of the CsI(Tl) to $\gamma$-rays}

After obtaining the $\gamma$ response spectra for each beam energy, both from simulations and experimental measurements, one can investigate the linear response of the CsI(Tl) to $\gamma$-rays. Fig.~\ref{LinearResults}  presents the energy deposit of the incident $\gamma$ obtained from the simulated spectra as a function of the ADC-E values obtained from the experimental measurement. Panels (a)-(d) denotes  the 4 different detector units tested in the experiment.  The red solid line corresponds to the linear fit, while the blue dashed curve represents the quadratic fit. As seen in all panels, the linear fit and the quadratic curve are very close to each other, indicating that all the CsI $\gamma$ detectors exhibit a good linear response to $\gamma$ energy below 20 MeV. 

To quantitatively evaluate the non-linearity in the $\gamma$ energy response of different detectors, we present the results of both linear and quadratic fits  in Table \ref{FittingCompare}. Here $k$ and $l$ are the slope and the intercept parameters of the linear fit,  respectively. The $p_0$, $p_1$ and $p_2$ are the three parameters in the quadratic fit. The two quantities are employed to evaluate the non-linearity of the CsI(Tl) response. (i) the relative uncertainty of the quadratic term coefficient  defined by $|\frac{\epsilon_2}{p_2}|$, which indicates the statistical significance of the quadratic term, and (ii) the relative contribution of the quadratic term, measured by the ratio of  $r_c=|\frac{p_2}{k}\text{CH}_m|$, where $\text{CH}_m$ represents the ADC-E channel position corresponding to 20 MeV in the linear fit of each CsI detector from the beam test. The results are shown in the last two columns in Table \ref{FittingCompare}. It is shown that the quadratic term coefficients for all four detectors are statistically insignificant, as the uncertainty $\epsilon_2$ of the quadratic term is at the same order of its value ${p_2}$. Their relative contributions of the quadratic term to the total energy  are negligible compared to the linear term, with $r_c$ ranging between 2.6\% to 4.5\% in accordance with the test results using the monochromatic $\gamma$-rays from proton-induced reactions \cite{Qin:2022mzp}. The result  confirms that the CsI(Tl) detectors exhibit excellent linearity in their $\gamma$ energy response. It is worth mentioning that unlike the case of measuring $\gamma$-rays,  the response of CsI(Tl) to light charged particle exhibits more pronounced non-linearity \cite{Guan:2021tbi,Wang:2021jgu}. The linearity of the CsI(Tl) detector response can be affected by several factors, most notably the spatial non-uniformity in optical photon collection. Differences between individual CsI crystals primarily arise from intrinsic variations including light coupling efficiency, the characteristics of the photomultiplier tubes (PMTs), and the properties of the associated readout electronics. These factors can vary significantly between detectors and should be carefully considered in the context of different experimental applications.

\section{A Calibration Scheme in Heavy Ion Experiments}

\begin{figure*}[hbtp]
    \centering
    \includegraphics[width=0.8\textwidth]{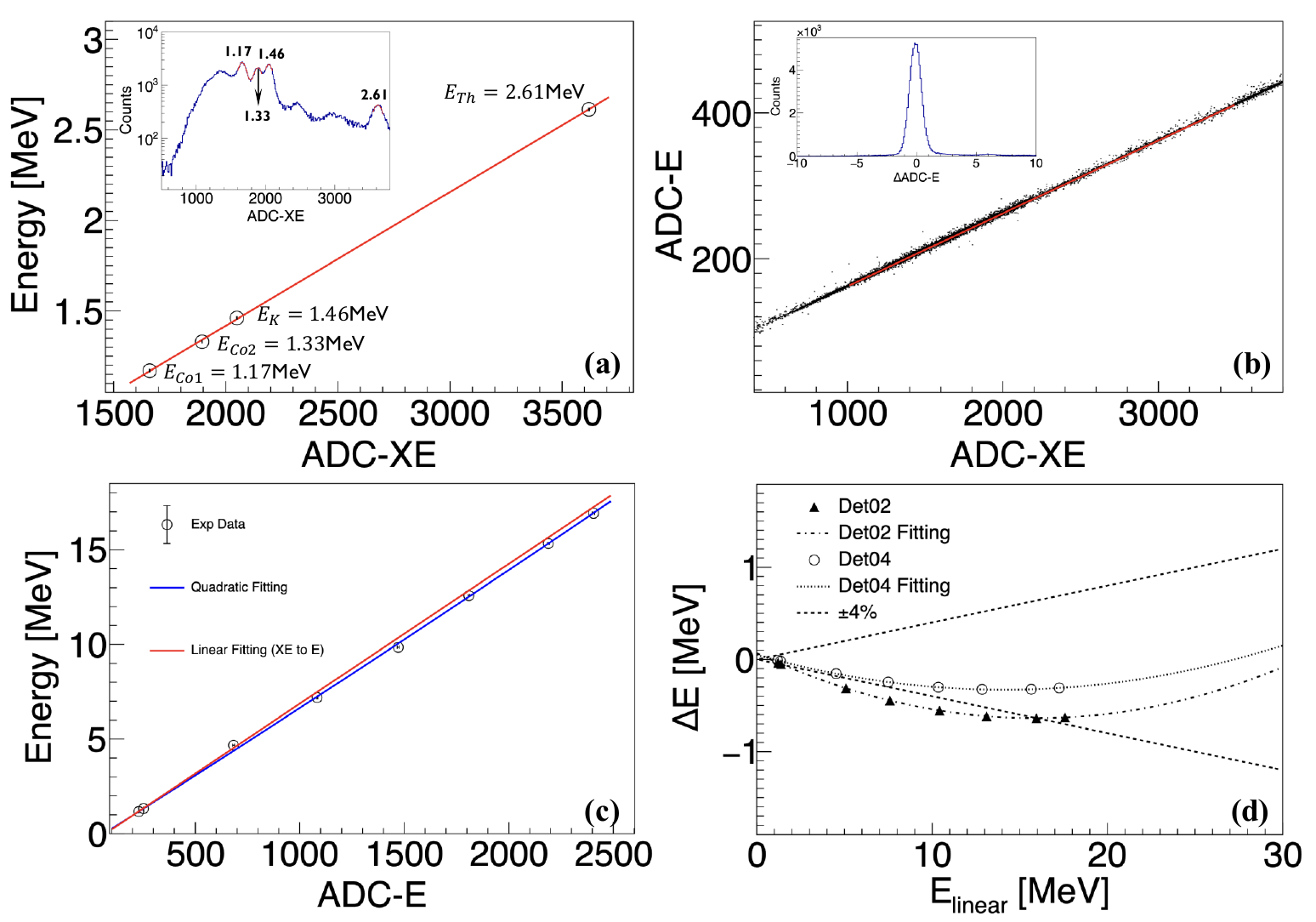}
    \caption{(Color Online) (a) Linear calibration of low-energy $\gamma$-rays with the ADC-XE output channel. (b) Relationship between the outputs of the main amplifier’s ADC-E and ADC-XE channels. (c) Nonlinear response of higher-energy $\gamma$-rays (blue curve) in comparison with the linear calibration result (red line) transformed from the ADC-XE to ADC-E. (d) Potential non-linear effects represented by $\Delta E$ between the blue curve and the red line in panel (c), see text for details.}
    \label{CSHINEapply}
\end{figure*}

Measuring the bremsstrahlung $\gamma$-rays produced in heavy-ion reactions at beam energies of tens of MeV/u is scientifically significant. These $\gamma$-rays carry information about the high-momentum tail of nucleons in the colliding system, offering insight into short-range nucleon correlations (SRCs) in atomic nuclei \cite{Qin:2023qcn, Xu:2024oct}. A few new experiments involving the high energy $\gamma$ hodoscope  have been conducted recently \cite{Qin:2023qcn, Si:2025eou, Xu:2025mvv} at CSHINE \cite{Guan:2021tbi, Si:2024ujh}. To measure the high-energy $\gamma$-rays, a $4\times4$ array of CsI crystals was installed. Details of the hodoscope can be found in \cite{Qin:2022mzp}. The energy deposited in each CsI(Tl) unit ranges from a few MeV to tens of MeV.

For calibration, high-energy  $\gamma$  sources are not available at the same accelerator facility. Typically, radioactive sources such as $^{60}$Co and $^{232}$Th are used to calibrate the detector’s linear response. These sources provide  $\gamma$  energies of 1.17 MeV, 1.33 MeV, and 2.61 MeV, which are much lower than the typical energy range used for the CsI(Tl) detectors in experiments. Extrapolating from these lower energies to the full range could introduce uncontrolled systematic errors. Therefore, developing a robust calibration scheme is crucial.

To extend the dynamic range of the detector and ensure accurate energy determination for both radioactive source  and high-energy $\gamma$ radiation, a double-range method is introduced. Specifically, we utilize the outputs from both the ADC-E and ADC-XE channels of the main amplifier, corresponding to low and high gain, respectively. The signals from the radioactive source, which are relatively small, are recorded by the ADC-XE output, while the energy of the bremsstrahlung $\gamma$-rays is read from the ADC-E channel. By using data from both channels, the calibration of the energy response can be achieved in the following steps.

\textbf{Step 1}: Calibration of the ADC-XE channel. The detector exhibits sufficiently good linearity in the $\gamma$ energy range of the source \cite{Qin:2022mzp}. Linear fitting is applied to build the relationship between the ADC-XE output signal at channel position $\rm {CH_{XE}}$  and the energy in MeV, which can be written as:
\begin{equation}
E_{\gamma} = a  \text{CH}_{\text{XE}} + b
\label{line_adc_xe}
\end{equation}

\textbf{Step 2}: Bridging the linear relationship between ADC-E and ADC-XE. This can be done by fitting the correlation of the ADC-E and the ADC-XE output in a wide range since every signals are fed to both channels and produce the self-correlation. The relation can be written as
\begin{equation}
    \text{CH}_{\text{E}} = \alpha \text{CH}_{\text{XE}}+\beta
\label{relation_xe_e}    
\end{equation}

\textbf{Step 3}: Generating the linear calibration of ADC-E. From  Eq.~(\ref{line_adc_xe}) and Eq.~(\ref{relation_xe_e}), it is feasible to  transform the linear calibration curve of the ADC-XE channel to the ADC-E channel and extrapolate to the entire range of the latter. Substituting Eq.~(\ref{line_adc_xe}) to Eq.~(\ref{relation_xe_e}), one obtains

\begin{equation}
    E_{\gamma} = \frac{a}{\alpha}\text{CH}_{\text{E}}+\frac{b\alpha-a\beta}{\alpha}.
    \label{line_adc_e}
\end{equation}

The scheme from step 1 to 3  allows us to achieve energy calibration for measuring higher-energy $\gamma$-rays.  It relies on the linearity of CsI(Tl) and the relation between the  ADC-E and ADC-XE channels of the main amplifier.

The validity of the calibration scheme mentioned above can be tested in the $\gamma$ beam experiment. Taking the unit Det04 for example in step 1, a $^{60}$Co radioactive source was placed near the CsI(Tl) detector but offset from the beamline to avoid obstruction. The ADC-XE channels covers well the  energy of the  $\gamma$-rays from the radioactive source and environmental background. Four peaks are visible on the raw spectrum, i.e., two peaks from $^{60}$Co source (1.17 MeV, 1.33 MeV), the $^{40}$K (1.46 MeV) peak and the $^{232}$Th peak (2.614 MeV) from the environment. The fitted peak ADC-XE positions were plotted alongside their corresponding energies in Fig.~\ref{CSHINEapply}(a) by the open circles, while the red line shows the linear fit. Excellent linearity of the response in the low energy range is demonstrated. The inset presents the energy spectrum of the ADC-XE channel. The distorted shape of the Compton distribution is attributed to environmental background, which does not affect the determination of the gamma peak positions of interest.

In step 2, Fig.~\ref{CSHINEapply}(b) presents the  the correlations between ADC-E and ADC-XE  channels for a large number of events.  The scattering dots represent the ADC-E and ADC-XE outputs for each event. A good linear relationship is seen between the two channels in the range of $1000 < {\rm CH_{XE}} < 3000$, out of which the linear correlation is severely distorted mainly because of the non-linearity of the ADC module. Using the intermediate region of $1000 < {\rm CH_{XE}}  < 3000$, the linear fitting is performed, as indicated by the red line. It builds the linear relationship of Eq.~(\ref{relation_xe_e}) between  ADC-E and ADC-XE outputs. To better illustrate the agreement between the data points and the fitted line, we show in the inset the residuals ($\Delta \text{ADC-E}$), defined as the difference between the measured ADC-E and the corresponding fitted value. The residuals form a narrow Gaussian distribution centered near zero, confirming the high accuracy of the linear correlation used in the analysis.

In Step 3, based on the linear transformation relationship between the ADC-XE and ADC-E channels shown in Fig.~\ref{CSHINEapply}(b), we can apply the linear calibration results from the ADC-XE channel (Fig.~\ref{CSHINEapply}(a)) to the ADC-E channel, as depicted by the red line in Fig.~\ref{CSHINEapply}(c). The open circles in panel (c) represent the analysis results of the $\gamma$  spectra obtained using the ADC-E channel. In addition to the six energy points from the  $\gamma$-ray
beam, two points, derived from fitting the $^{60}$Co source spectrum in the ADC-E channel, are also depicted. A quadratic fit was applied to these eight points, with the resulting curve shown as the blue solid line. It is clear that these eight energy points from the ADC-E channel align well with the linear response transformed from the low-energy  $\gamma$  calibration of the ADC-XE channel (red solid line), and the difference between the red solid line (from the low-energy calibration) and the blue curve (from the quadratic fit) is minimal.

The difference $\Delta E = E_{\text{quadr}} - E_{\text{lin}}$  between the red line and the blue curve in Fig.~\ref{CSHINEapply}(c)  provides a quantitative measure of the non-linearity of the CsI(Tl) response  to high-energy $\gamma$-rays and can be considered as the systematic  uncertainty of the linear calibration scheme. Fig.~\ref{CSHINEapply}(d) presents the distribution of  $\Delta E$  as a function of $E_\gamma$ obtained by the linear calibration. The open circles are the experimental data points.  Quantitatively, it is shown that the deviation in $\gamma$  energy calibration around 20 MeV using this method is approximately 2\%. These results confirm the high accuracy of this experimental calibration scheme for $\gamma$-rays in the 20 MeV range. In the case of Det02 detector, which exhibits larger non-linearity, the deviation (solid triangles) increases to around 4\% at 20 MeV.  It suggests that the systematic uncertainty introduced by the linear calibration method remains at the similar level of the amplitude of non-linearity, as listed in Table~\ref{FittingCompare}. The quantitative results are consistent with our previous studies \cite{Qin:2022mzp}, where proton-induced reactions on a LiF target were used to produce monochromatic $\gamma$-rays.

Finally, the non-linearity can be translated into the systematic uncertainty of the energy spectrum of the bremsstrahlung $\gamma$-rays in the HIC experiment at CSHINE. Specifically, by applying different $\Delta E$  modifications (2\% and 4\%) to the energy response, various $\gamma$-ray energy spectra can be generated. The variance of these spectra can then be calculated bin-by-bin as the systematic uncertainties, prior to conducting the subsequent physical analyses \cite{Qin:2023qcn,Xu:2024oct}. The same scheme will be applied in the upcoming 25 MeV/u $^{124}$Sn+$^{124}$Sn experiment \cite{Si:2024ujh}.

\section{Conclusion}

In this study, we present a comprehensive calibration and characterization of the CsI(Tl) $\gamma$-ray detector, which is crucial for measuring high-energy $\gamma$-rays produced in heavy-ion collision (HIC) experiments. Through extensive beam experiments at the SLEGS beamline station, we successfully obtained energy response spectra for various $\gamma$ energies. The calibration process, combining experimental measurements and Geant4 simulations, demonstrates that the CsI(Tl) detector exhibits a good linear response to $\gamma$ energies up to 20 MeV. The non-linearity is at the level of 4\%.

We also highlight the accuracy and feasibility of the calibration scheme  applied to the CSHINE experiment, where the CsI crystal array functions as an electromagnetic calorimeter for high-energy $\gamma$-rays produced in HICs. The results reveal that the deviation of a linear fit from the quadratic fit in $\gamma$ energy calibration around 20 MeV is below 4\%, which can be considered as a systematic uncertainty source. This calibration is vital for studying short-range nucleon correlations in atomic nuclei, as the reconstructed high-energy $\gamma$ spectra provide valuable insights into the high-momentum tail of nucleon momentum distributions. These results lay a solid foundation for the future application of CsI(Tl) in high-energy  $\gamma$-spectroscopy research in nuclear physics, particularly within the context of HIC experiments.

{\it Acknowledgements.} The authors gratefully acknowledge the support from the National Natural Science Foundation of China (Grant Nos. 12335008 and 12205160) and Tsinghua University Initiative Scientific Research Program.

\end{document}